\documentstyle[aps,%
prc,%
twocolumn,%
floats,%
epsfig%
]{revtex}

\tighten

\newcommand{\half}[1]{{\small\frac{#1}{2}}}
\newcommand{\Mhat}{{\hat{\cal M}}}
\newcommand{\nuc}[2]{{{$^{#1}$#2}}}
\newcommand{\nil}[2]{{$[#1]\half{#2}$}}

\newcommand{\kjd}{{k_{J_d}}}

\newcommand{\Imag}{{\cal I}{\rm m}}
\newcommand{\Reg}{{{\cal R}eg}}

\begin{document}
\twocolumn[\columnwidth\textwidth\csname@twocolumnfalse\endcsname

\title{Theoretical description of deformed proton emitters:\\  
nonadiabatic coupled-channel method}
\draft
\author{
  B. Barmore,$^{1-3}$,
  A.T. Kruppa,$^{2,4}$,
  W. Nazarewicz,$^{1,3,5}$ and
  T. Vertse$^{2,4}$ }

\address{$^{1}$Department of Physics and Astronomy, University of Tennessee,
  Knoxville, Tennessee 37996}
\address{$^{2}$Joint Institute for Heavy Ion Research, Oak Ridge
  National Laboratory, P.O. Box 2008, Oak Ridge, Tennessee 37831}
\address{$^{3}$Physics Division, Oak Ridge National Laboratory, P.O.
  Box 2008, Oak Ridge, Tennessee 37831}
\address{$^{4}$Institute of Nuclear Research of the Hungarian
  Academy of Sciences, P.O. Box 51, H-4001, Debrecen, Hungary}
\address{$^{5}$Institute of Theoretical Physics, Warsaw University,
  ul. Ho\.{z}a 69, PL-00681, Warsaw, Poland}

\maketitle
\begin{abstract}
The newly developed nonadiabatic method based on the coupled-channel 
Schr\"odinger equation with Gamow states is
used to study the phenomenon of proton radioactivity. 
The new method, adopting the
weak coupling regime of the particle-plus-rotor model, allows for the
inclusion of excitations in the daughter nucleus.  This can lead to
rather different predictions for lifetimes and branching ratios as
compared to the standard adiabatic approximation corresponding
to the strong coupling scheme.  Calculations are
performed for several experimentally seen, non-spherical nuclei
beyond the proton dripline.  By comparing theory and experiment, we
are able to characterize the angular momentum content of the
observed narrow resonance.
\end{abstract}


\addvspace{5mm}]  

\narrowtext

\section{Introduction}
Nuclei  beyond the proton dripline are unstable against proton
emission. Although formally unbound, some of these systems have 
rather long lifetimes, ranging  from microseconds to  seconds,
due to the confining effect of the Coulomb  barrier
\cite{[Hof95b],[Woo97]}.

The past few years have seen an explosion of exciting discoveries in
this field including new ground-state and isomeric proton 
emitters~\cite{[Dav98],[Bat98],[Ryk99],[Bin99]} and the
first evidence for fine structure in proton decay~\cite{[Son99]}. 
The focus of recent investigations has been on
well-deformed nuclei which 
exhibit collective motion.  These are of particular interest due to
the interplay between proton emission and angular momentum.  

The theoretical description of long-lived proton emitters requires a
detailed understanding of narrow resonances. 
Although proton radioactivity is a complicated $A$-body phenomenon, much
insight may be gained by considering the simplified problem of a
single proton penetrating the Coulomb barrier of the core consisting
of the remaining $A$--1 nucleons.
It has been found that this simple one-body picture works
surprisingly well.  In many cases one  has been able to determine the
angular momentum content of the resonance and the associated
spectroscopic factor \cite{[Woo97]}.

For spherical systems, there are many  methods on the market which
give similarly precise descriptions and, 
in many cases, one  has been able to determine the
angular momentum content of the resonance and the associated
spectroscopic factor \cite{[Woo97],[Abe97]}. 

The array of theoretical tools available 
for deformed emitters is not as well developed.  The existing ones
fall into three general categories.  The first 
family of calculations~\cite{[Dav98],[Son99],[Bug89]} is based on the
reaction-theoretical framework of Kadmenski\u{\i} and
collaborators \cite{[Kad73]}.  The
second suite uses the theory of Gamow (resonance)
states~\cite{[Ryk99],[Fer97],[Mag98],[Mag99]}.  Finally,
an approach, based on the time-dependent Schr\"{o}dinger equation,
has been introduced in Ref.~\cite{[Tal98]}.    

In all of these previous attempts, the strong coupling approximation of 
the particle-plus-rotor model has been used.  The core is taken to be
a perfect rotor with infinite moment of inertia.  This has the effect
of (i) collapsing the rotational spectrum of the daughter nucleus
to the ground
state and (ii) neglecting the 
Coriolis coupling.  Recently we have introduced a 
technique based on the weak coupling scheme which is free from these 
deficiencies \cite{[Kru00]}.
  Within this method, partial proton widths from different 
states of the parent nucleus to various final states in the daughter system can
be calculated in a straightforward and consistent manner.

We will begin in Sec.~\ref{sec:theory} by laying the theoretical
framework for this work.  Section~\ref{sec:numeric} discusses the
 numerical methods adopted in our work.
 Section \ref{sec:phys}  presents
application of the method to the structure of deformed proton emitters.
   A critical 
analysis  of the adiabatic and nonadiabatic methods is contained in
Sec.~\ref{sec:adi}. Finally, conclusions are given in
Sec.~\ref{sec:conc}.


\section{Theoretical Basis} \label{sec:theory}

From a theoretical point of view, proton radioactivity is an excellent
example of three-dimensional, quantum-mechanical tunneling.
As such, the understanding of proton emission is really a test of our
knowledge of very narrow resonances.  Since the lifetimes which can be
seen experimentally range from microseconds to seconds,  
the corresponding widths are extremely small; they vary
 between $10^{-16}$ MeV and $10^{-22}$ MeV\@.  
Theoretical description of such small widths requires
high numerical accuracy. In the following,  
 the coupled-channel Schr\"{o}dinger equation method with Gamow
 states is outlined, and the proton-plus-core Hamiltonian is defined.

\subsection{Coupled-channel Equations}

\begin{figure}[tbp]
  \begin{center}
 \leavevmode
    \epsfig{file=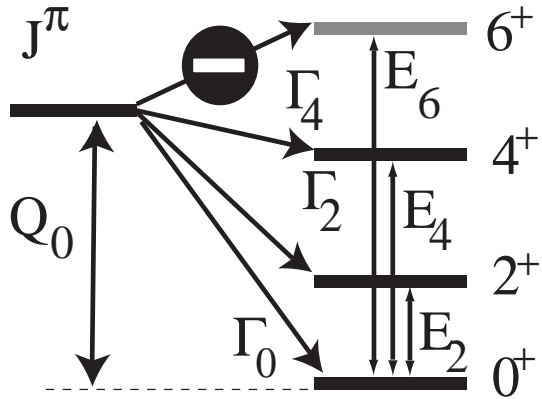,width=8cm}
  \end{center}
  \caption{Schematic diagram illustrating 
    the energetics of proton decay.  A $J^\pi$ state of an odd-$Z$
    parent nucleus (left) decays to those members of the ground-state
    rotational band of the even-even deformed daughter nucleus (right)
    which are in the $Q$ window. The band members have excitation
    energies $E_{J_d}$ relative to the ground state, and $Q_0$ is the
    $Q_p$ value for the decay to the ground state.  As shown here,
    usually only a few channels are energetically open. The
    corresponding partial widths are indicated by $\Gamma_{J_d}$.  }
  \label{fig:defs}
\end{figure}

The parent nucleus  is described by the core-plus-proton
Hamiltonian,
\begin{equation}\label{Htot}
H=H_{d} + H_p + V,
\end{equation}
where $H_d$ is the Hamiltonian of the daughter nucleus, $H_p$ is that of the
proton, and $V$ is the proton-daughter interaction.  In the weak coupling
scheme, the 
wave function of the parent nucleus is written as 
\begin{equation}\label{Psi}
\Psi_{JM}=r^{-1}\sum_{J_dl_pj_p} u_{J_dl_pj_p}^J(r)
\left({\cal Y}_{l_pj_p}
\otimes \Phi_{J_d}\right)_{JM}.
\end{equation}
This wave function is labeled by parity, total angular momentum $J$, and its
projection $M$.  In Eq.~(\ref{Psi}), $u_{\alpha}^J(r)$ [where
$\alpha \equiv (J_dl_pj_p)$ completely labels the channel quantum
numbers] 
is the cluster radial wave function representing the relative radial
motion of the proton and the core,  and ${\cal Y}_{l_pj_pm_p}$ is
the orbital-spin wave function of the proton.  The daughter wave function,
$\Phi_{J_dM_d}$, satisfies 
\begin{equation}\label{HPsi}
H_{d} \Phi_{J_dM_d} = E_{J_d} \Phi_{J_dM_d}.
\end{equation}
In the present formalism,  the daughter's spectrum does not have
to be known explicitly.
Where possible, the energies  $E_{J_d}$ are  taken from experiment;
otherwise,  the spectrum is modeled theoretically.  Figure~\ref{fig:defs} shows a 
schematic diagram illustrating the energetics of proton emission from 
a $J^\pi$ state of an
odd-$Z$ parent nucleus to the ground-state rotational band of
the deformed daughter nucleus.

As usual, the coupled-channel equations are obtained by 
inserting Eq.~(\ref{Psi}) into the Schr\"{o}dinger equation and integrating over all 
coordinates, save the radial variable $r$ \cite{[Bug89],[Tam65]}:
\begin{eqnarray}\label{cc}
  \left[-\frac{\hbar^2}{2\mu} \frac{d^2}{dr^2} \right.
& + & \left. \frac{\hbar^2 l_p(l_p+1)}{2\mu r^2} +
  V_{\alpha\alpha}(r) - Q_{J_d} \right]  u_{\alpha}^J(r) \nonumber\\
 & + & \sum_{\alpha'\ne\alpha} V_{\alpha,\alpha'}^J(r) \;
 u_{\alpha'}^J(r)=0.
\end{eqnarray}
In Eq.~(\ref{cc}), $V_{\alpha\alpha}$ is the diagonal part of the
proton-core potential, $Q_{J_d}$ is the energy of the emitted proton
leaving the daughter nucleus in the state $J_d$, and $V^{J}_{\alpha
  \alpha'}$ are the off-diagonal coupling terms.  The $Q_{J_d}$ values
follow from the spectrum of the daughter nucleus, $Q_{J_d} = Q_0 -
E_{J_d}$, where $Q_0$ is the $Q_p$ value for the decay to the $0^+$
ground state (see Fig.~\ref{fig:defs}).

To illuminate the dynamics of the system, one  can expand the
proton-daughter potential in multipoles~\cite{[Tam65]},
\begin{equation}\label{eq:multipole}
  V = \sum_{\lambda} v_{\lambda}(r) \, (\Mhat_{\lambda} \otimes Y_{\lambda})_{00}.
\end{equation}
The matrix elements $V_{\alpha,\alpha'}^J(r)$ can then be written in
the simple, yet generic, form
\begin{equation}\label{Vaa'}
  V_{\alpha,\alpha'}^J(r) = \sum_{\lambda} v_{\lambda}(r) \; 
  \langle J_d || \Mhat_{\lambda} || J'_d \rangle \;
  {\cal A}(l_p j_p J_d,\, l'_p j'_p J'_d, \,\lambda J).
\end{equation}
The factor $\cal A$ is purely geometric and comes from the proper
coupling of angular momentum vectors.  The reduced matrix elements of
$\Mhat_{\lambda}$ contain all of the dynamics of the core.  
Since we consider only rotational nuclei in this paper, they are given by a simple
expression~\cite{[Tam65]}
\begin{equation}
  \langle J_d || \Mhat_{\lambda} || J'_d \rangle = \sqrt{2 J'_d + 1} \;
  \langle J'_d \lambda \; K 0 | J_d K \rangle.
\end{equation}
To consider other excitation modes in the daughter system, one needs only 
change these reduced matrix elements.

To be a resonant state, the cluster radial wave function must vanish
at the origin and behave as an outgoing Coulomb wave, $O_l=G_l+iF_l$, 
beyond the range of the nuclear interaction and the off-diagonal Coulomb interaction,
\begin{eqnarray}\label{asympwf}
  u^{J}_{J_d l_p j_p}(r) &\stackrel{{\rm large}\; r}{\longrightarrow}
  &O_{l_p}(\eta_{J_d},r\kjd)\nonumber \\
  &=&G_{l_p}(\eta_{J_d},r\kjd) + i F_{l_p}(\eta_{J_d},r \kjd) ,
\end{eqnarray}
where $\kjd^2=2\mu Q_{J_d}/\hbar^2$ and
$\eta_{J_d}\kjd=\mu Ze^2/\hbar^2$.
These two conditions are only satisfied for a discrete set of 
complex wave numbers $k$.
The generalized eigenvalues of Eq.~(\ref{cc}) correspond to the poles
 of the scattering
matrix~\cite{[Hum61],[Ver87]}.
The corresponding solutions are either bound or antibound
states, ${\cal E}=E_b<0$, with negative real energies and
imaginary wave numbers $k=i\gamma$ ($\gamma>0$ for bound and
$\gamma<0$ for antibound states ), or resonance states,
${\cal E}=Q -i {\Gamma\over 2}$, with a nonzero imaginary part
$\Gamma \neq 0$, and $k=\kappa-i\gamma$.
 
\begin{figure}[tbp]
  \begin{center}
\leavevmode
    \epsfig{file=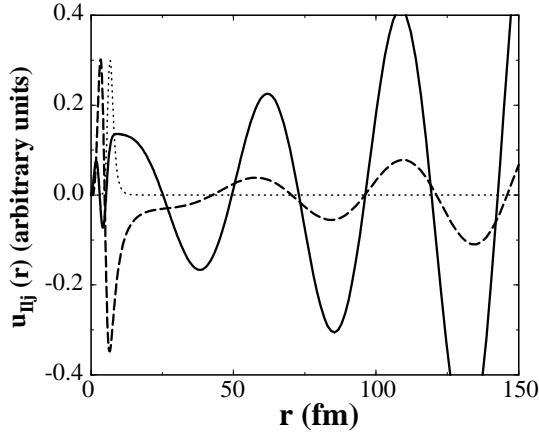,width=8cm}
  \end{center}  
    \caption{Asymptotic
      behavior of three-channel wave functions corresponding to 
      the $J^\pi$=$3/2^-$ neutron resonance in \nuc{141}{Ho} at energy  
      ${\cal E}=(0.378 -i 0.0732$)\,MeV calculated at  $\beta_2$=0.20.
      The solid line denotes  the $p_{3/2}\otimes 0^{+}$ channel.  Notice the
      increasing envelope for increasing $r$.  The dashed line labels  the 
      $f_{5/2}\otimes 2^{+}$ channel function.  The dotted line 
      corresponds to  the closed channel
      $l_{19/2}\otimes 8^{+}$.  The wave function decays exponentially 
      since  $Q<0$.
      The magnitude has been scaled so all 
      three wave functions could be shown.}
    \label{fig:wavefunction}
\end{figure}

The asymptotic behavior of these solutions is determined by $k$; at a
very large distance the outgoing solution is proportional to $e^{i k
  r}$.  For resonance states, $e^{i k r}=e^{i\kappa r}e^{\gamma r}$,
i.e., the wave function {\rm diverges} exponentially.  As discussed in
Refs.~\cite{[Hum61],[Ver87]}, this seemingly unphysical feature of
Gamow wave functions has a natural explanation in the fact that Gamow
states do not represent time-dependent wave packets but static
sources.  To illustrate the asymptotic behavior of Gamow wave
functions, Fig.~\ref{fig:wavefunction} shows three-channel wave
functions corresponding to a broad neutron resonance.

Due to the divergent  behavior at large $r$, one  must define a new
normalization scheme for the Gamow states.  
Berggren proposed a new completeness relation, which includes Gamow
states~\cite{[Ber68]}, by generalizing the scalar product. He
introduced  a bilinear basis set and a regularization
procedure ($\Reg$).  With this generalization, the norm is
\begin{equation}
\label{sphnorm}
\sum_{\alpha} \Reg \int_0^{\infty} [u^J_{\alpha}(r)]^2 \, dr 
= 1 .
\end{equation}
A convenient method for regularization is  to rotate
 $r$ into the first quadrant of the complex $r$-plane
beyond a certain distance $r_{\rm max}$. This is often referred to as
the exterior complex scaling method. (For other regularization techniques,
see Ref.~\cite{[Ver87]}.)

Once we know the resonance energy and radial wave functions, there are
several methods to calculate the width of the state.  The most
straightforward method is to take twice the negative of the imaginary
part of the resonance energy.  However, for the narrow resonances
associated with proton emitters, the numerical accuracy needed to
calculate $\Imag [{\cal E}]$ is difficult to achieve.  Therefore,
approximate methods are often used.
 
One possibility is to calculate the partial width for each channel
from the current expression~\cite{[Hum61]},
\begin{equation}
\label{partcur}
\Gamma_{\alpha}(r)=i\frac{\hbar^2}{2\mu}  
\frac{u_{\alpha}^{\prime*}(r) u_{\alpha}(r) 
- u^{\prime}_{\alpha}(r) u_{\alpha}^*(r)}
{\sum_{\alpha^{\prime}} \int_0^{r} |u_{\alpha^{\prime}}(r')|^2 
  {\rm d}r'} ,
\end{equation}
where the sum of the partial widths,
\begin{equation}
\label{totgam}
\Gamma (r) = \sum_{\alpha} \Gamma_{\alpha}(r) ,
\end{equation}
gives the total decay width (see Fig.~\ref{fig:defs}).  Although
values of $\Gamma_{\alpha}(r)$ depend on $r$ in the region
$r$$<$$r_{\rm as}$ where the coupling potential terms are not
negligible, $\Gamma(r)$ is strictly independent of $r$ by construction
($\Gamma (r)=-2\Imag [{\cal E}]=\Gamma$) which reflects the flux
conservation (continuity equation).  Beyond the asymptotic radius,
$r_{\rm as}$, the partial widths, $\Gamma_{\alpha}(r)$ have a
negligible dependence on radius.  We take $r_{\rm as} \approx 40$fm.
Numerically, $\Gamma (r)$ varies little with distance and differs by
less than $0.1\%$ from the $\Gamma$ obtained from the imaginary part
of the eigenvalue.

The Gamow boundary condition given by Eq.~(\ref{asympwf})  
is usually written in the form 
\begin{equation}\label{gbcond}
\frac {u'_\alpha (r_{\rm as})}{u_\alpha (r_{\rm as})}=\kjd\frac{ 
O'_{l_p}(\eta_{J_d},r_{\rm as}\kjd)}{O_{l_p}(\eta_{J_d},r_{\rm as}\kjd)},  
\end{equation}
where $r_{\rm as}$ is the channel radius.  (The off-diagonal couplings 
are negligible beyond it.) Using  Eq.~(\ref{gbcond}),
the partial decay widths can be
written at the point $r_{\rm as}$ as  
\begin{eqnarray}
\label{gwidth}
\Gamma_{\alpha}(r_{\rm as})& = &i\frac{\hbar^2 }{2\mu}  
\frac{|u_{\alpha} (r_{\rm as})|^2}
{|O_{l_p}(\eta_{J_d}, \kjd r_{\rm as})|^2 
\sum_{\alpha^{\prime}} \int_0^{r_{\rm as}}
 |u_{\alpha^{\prime}}(r')|^2 
  {\rm d}r'}\nonumber\\
  & \times & \left[ k_{J_d}^{*} O'^*_{l_p}(\eta_{J_d},r_{\rm as}\kjd) 
  O_{l_p}(\eta_{J_d},r_{\rm as}\kjd) \right. \nonumber\\
 & - & \left. \kjd O'_{l_p}(\eta_{J_d},r_{\rm as}\kjd) 
  O^*_{l_p}(\eta_{J_d},r_{\rm as}\kjd) \right].
\end{eqnarray}
If we neglect the very small imaginary part of $\kjd$, the square bracket 
in Eq.~(\ref{gwidth}) is equal to 
$-2i$.  Furthermore, if we  assume that  for a very narrow resonance the
imaginary part of $u_{\alpha}$ is very small
(hence the generalized normalization 
condition (\ref{sphnorm}) is roughly equivalent to the ``normal"
normalization 
$\sum_{\alpha^{\prime}} \int_0^{r_{\rm as}} |u_{\alpha^{\prime}}(r')|^2 
dr'\approx 1$), 
 then we end up with the approximate expression
 for the partial decay width:
\begin{equation}\label{gwidth2} 
\Gamma_{\alpha}(r_{\rm as}) \approx \frac{\hbar^2 \kappa_{J_d}}{\mu}  \frac{ 
|u_{\alpha} (r_{\rm as})|^2}
{|O_{l_p}(\eta_{J_d}, \kjd r_{\rm as})|^2}.
\end{equation}
It is to be noted that 
Eq.~(\ref{gwidth}) and its approximate form (\ref{gwidth2}) are valid only
at the point $r_{\rm as}$. The expression (\ref{gwidth2}) was used in
papers \cite{[Mag98],[Mag99],[Fer00],[Mag00]}.  We emphasize that
if the coupled equations are solved with the Gamow boundary condition,
then the total width can be calculated at any intermediate point
using Eqs.~(\ref{partcur}) and (\ref{totgam}).  The expression
(\ref{gwidth2}) is very similar to that of the R-matrix
theory (see below), but it relies on different approximations and boundary
conditions than the R-matrix formalism.

In the R-matrix theory, we also have a set of radial functions,
$g_{\alpha}$.  These functions are regular at the origin and satisfy
the coupled equations but with the following boundary
conditions:
\begin{equation}\label{rbcond}
r_{\rm as} \frac {g'_\alpha (r_{\rm as})}{g_\alpha (r_{\rm as})}=B_\alpha,    
\end{equation}
where $B_\alpha$ are arbitrary real numbers. Due to the real boundary
condition, the R-matrix eigenvalues are real numbers.  In the R-matrix
theory, the wave function is normalized inside the sphere of radius
$r_{\rm as}$, i.e., $\sum_\alpha \int_0^{r_{\rm as}} |g_\alpha(r)|^2
dr=1$.  Thomas has shown \cite{[Tho54]} that in a one-level
approximation with appropriately chosen $B_{\alpha}$, in which the
level shift is ignored, the position of the Gamow resonance
corresponds to the R-matrix eigenvalue, and the width of the state
can be calculated in the form given by Eq.~(\ref{gwidth2}) in which
$u_\alpha (r)$ is replaced with $g_\alpha (r)$.  This R-matrix
approximation works fairly well \cite{[Ari74],[Kru99]} for very narrow
Gamow resonances corresponding to the known proton emitters. For large
values of the channel radius $r_{\rm as}$, expression (\ref{gwidth2})
is generally within 2\% of the values calculated explicitly from
Eq.~(\ref{cc}) or obtained via the current expression (\ref{partcur}).
a detailed comparison of the R-matrix theory and the Gamow formalism for
proton emitters will be given in Ref.~\cite{[Kru00a]}.

The nonadiabatic approach allows for a straightforward
calculation of branching ratios.  The partial width 
corresponding to the decay to a core
state $J_d$ is given by 
\begin{equation}
  \Gamma_{J_d} = \sum_{\{lj\}} \Gamma_{{J_d}lj},
\end{equation}
where $\Gamma_{{J_d}lj} = \Gamma_{\alpha}$ is 
 given by Eq.~(\ref{partcur}).  
Once the total width is known, the half-life  for  proton emission is  
\begin{equation}
  T_{\half{1}} = \frac{\hbar \ln 2}{\Gamma} .
\end{equation}

The use of the weak coupling scheme represented by Eq.~(\ref{Psi}) has
several advantages.  First, excitations of the core are included in a
straightforward manner.  This enables us to study the proton decay
from the rotational bands of the parent nucleus to the ground-state
rotational band of the daughter nucleus.  Furthermore, since the
formalism is based on the laboratory-system description [Hamiltonian
(\ref{Htot}) is rotationally invariant and the wave function
$\Psi_{JM}$ conserves angular momentum], the Coriolis coupling is
automatically included.

\subsection{Strong Coupling Limit}\label{sec:strong}
A great simplification to Eq.~(\ref{cc}) occurs if one considers
all of the rotational states in the daughter's ground-state band to be 
degenerate (i.e., $Q_{J_d} \equiv Q_p$ for all $J_d$).  This is the
limit of  strong coupling where the moment of inertia of the
daughter is taken to
infinity.  It is also the {\it adiabatic approximation} of
Refs.~\cite{[Tam65],[Bar64]}. 

In this limit, the coupled-channel equations, (\ref{cc}), reduce
to those for the intrinsic (i.e., deformed)
Nilsson orbital~\cite{[Bug89]}
\begin{equation}\label{Nills}
  \Psi_{\Omega}=\sum_{j_pl_p} \frac{u_{\Omega\Omega j_pl_p}(r)}{r}{\cal
    Y}_{l_pj_p\Omega}
\end{equation}
where
\begin{equation}\label{adiabatic}
   u_{JK\,j_pl_p} = \sqrt{2} \ (-1)^{K+J}\sum_{J_{d}}
\langle J_d 0 j_p K | JK \rangle u^J_{J_{d}l_pj_p}.
\end{equation}
In Eq.~(\ref{Nills}) $\Omega$=$K$=$J$ is the angular momentum
projection on the symmetry axis.  As seen from Eq.~(\ref{adiabatic}),
the strongly coupled intrinsic state contains contributions from all
the cluster wave functions corresponding to {\em different core
  states.}  Since, as discussed by Tamura\cite{[Tam65]}, there is no
dynamic coupling between the angular momentum of the proton and that
of the daughter nucleus (the daughter nucleus is perfectly inert
during the proton emission), there exist infinitely many solutions
obtained by combining $\bbox{j}_p$ and $\bbox{J}_d$.  Since the core
states are degenerate, all the solutions with $J\ge\Omega$ are
degenerate as well.

\subsection{Model Parameters}\label{sec:mod-para}

\widetext
\begin{figure*}[hbt]
  \begin{center}
\leavevmode
    \epsfig{file=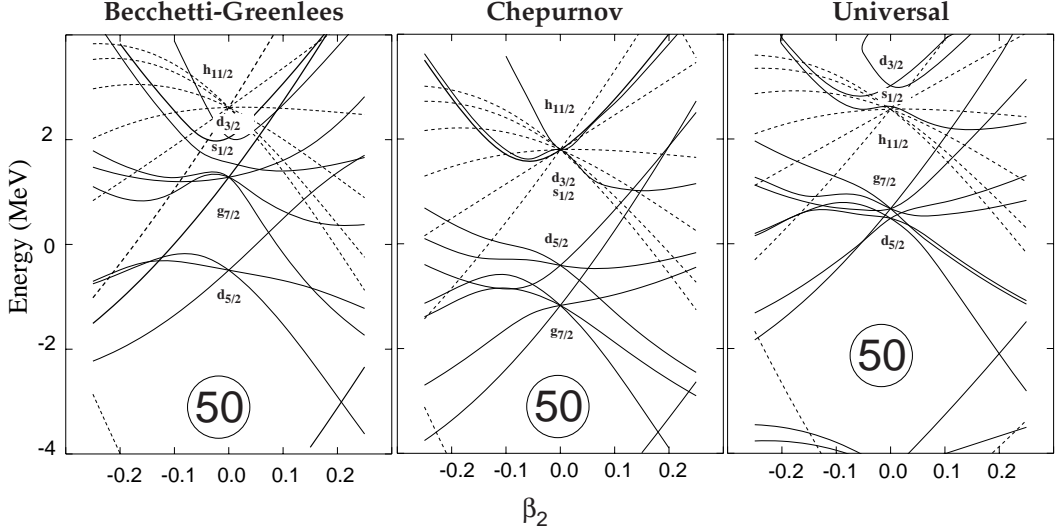, width=15cm}
  \end{center}
\vspace*{-4cm}
    \caption{Comparison of deformed 
    single-proton  levels for $Z=63, N=68$ predicted in
    three WS parameterizations.
   The left panel
      shows the Nilsson diagram calculated with the
      Becchetti-Greenlees set which yields  poor  ordering of the 
      single-particle levels but good radial properties.  The right
      panel is obtained with the  ``universal" set which yields good ordering 
      of Nilsson levels but poor
      radial properties of wave functions.  The center panel 
      was obtained with the  Chepurnov parameter set used
      in this work. This latter parameterization gives a very reasonable compromise
      between the radial and spectroscopic properties.}
      \label{fig:splevels}
\end{figure*}
\narrowtext

In this work, we assume that the average single-particle potential is
approximated by the sum of a Woods-Saxon (WS) potential, a spin-orbit
term, and a Coulomb potential. The axially deformed WS potential is
defined according to Ref.~\cite{[Cwi87]}. We employ the Chepurnov
parameterization \cite{[Che67a]}; it is in 
good agreement with the proton
single-particle energy levels given in  the systematic study \cite{[Naz90]}.  
{\AA}berg et al.~\cite{[Abe97]} discussed the effect of the optical
model parameters on spherical proton emitters.  They concluded that
the uncertainties in the parameters affect the half-lives by, at most, a 
factor of 3.  For spherical proton emitters, they concluded that the
Becchetti-Greenlees WS potential \cite{[Bec69]}, commonly used in 
spherical calculations for proton emitters,  was better than
the universal
parameter set \cite{[Dud81]} (excellent for the description of
structure properties of deformed rare-earth nuclei \cite{[Naz90]} but
having too large a radius to give a quantitative description of the
tunneling rate). Since, for the description
of spherical proton emitters, the nodal behavior of radial wave functions 
plays a minor role \cite{[Abe97]}, the actual order of 
spherical shells does not really matter.

However, in the case of  deformed proton emitters the situation is  different.
While the radial properties of the optical model potential are still
important, the proper ordering of
spherical shells becomes crucial since it affects the fragmentation
of orbital angular momentum caused by  deformation. In this context,
as illustrated in Fig.~\ref{fig:splevels},
the Becchetti-Greenlees parameter set  performs  rather poorly, while
the Chepurnov parameterization offers a compromise between  good
radial properties  and proper level ordering.

Since within any mean-field theory the resonance energy
cannot be predicted with sufficient accuracy,
following Refs.~\cite{[Ryk99],[Abe97]}, the depth of the WS potential
is adjusted to give the experimental $Q_0$ value.  The deformed part
of the spin-orbit interaction is neglected; we do not expect this to
have a noticeable effect on the results~\cite{[Nil55]}.  The
off-diagonal coupling in (\ref{cc}) appears thanks to the
non-spherical parts of the WS and Coulomb potentials.

\begin{figure}[hbt]
  \begin{center}
\leavevmode
    \epsfig{file=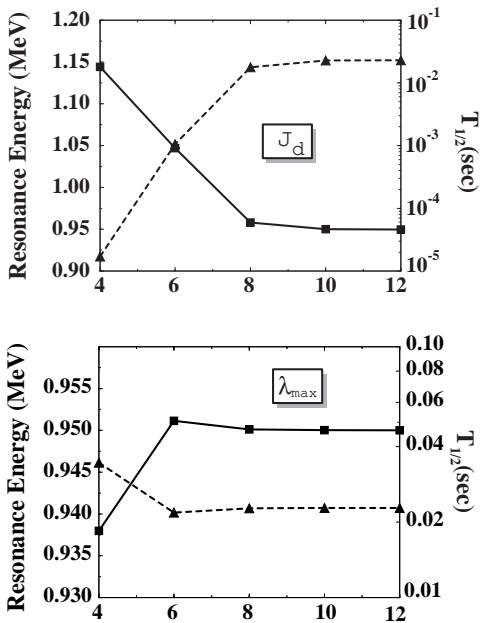,width=8cm}
  \end{center}
\vspace*{-1cm}
    \caption{Dependence of the resonance eigenstate on various expansion 
      parameters.  Calculations were done for the \nil{411}{3} level in
      \nuc{131}{Eu} at a deformation of $\beta_2 = 0.37$.  The upper
      panel shows the resonance energy (solid line with squares and
      left scale) and the lifetime (dashed line with triangles and
      right scale) as a function of the number of included states in
      the ground-state band of the daughter nucleus.  The lower panel
      shows the same except as a function of the number of spherical
      multipoles used in expanding the deformed single-particle potential.}
    \label{fig:conv}  
\end{figure}

Great care was taken to ensure that enough channels were considered in 
solving Eq.~(\ref{cc}) for proper convergence in the
eigenvalues.   As seen in the lower panel of Fig.~\ref{fig:conv}, 
expanding the WS in spherical multipoles to order 8 is sufficient for
convergence.  However, to be on the safe side,
a value of $\lambda_{\rm max} = 12$ in
 Eq.~(\ref{eq:multipole}) was used in all calculations.  The number 
of partial waves that were needed in the decomposition of the proton
radial wave function varies from system to system
depending mainly on the angular momentum of the proton state. In
general, all partial waves with $l $$< $10--13 are needed.
Since, in the nonadiabatic approach, the maximum proton
angular momentum  
and the maximum daughter spin considered are closely related, the above
condition corresponds to $(J_d)_{\rm max} = 10$ which was used for
all calculations (see upper panel of Fig.~\ref{fig:conv}).  
Since the high-spin channels are energetically
forbidden, their exact placement is of minor importance.  Only
the energy of the $2^+$ level and, occasionally, the $4^+$ level have a
profound effect on the resonance energy and other observable
quantities.  (For more discussion concerning this point, see
Sec.~\ref{sec:adi}.)

It is important in this work that we have a good representation of the
ground-state rotational band in the daughter nucleus.  In a few of the
systems studied in this work, such spectra exists for $J_d \le 10$,
which is enough for adequate convergence.  However, for the most
highly deformed systems, the spectroscopic information does not exist.
For these nuclei, we parameterize the ground-state rotational band as
$E_{J_d}$= $\kappa J_d (J_d+1)$, where $\kappa$ is adjusted to the
$E_{2^+}$ energy.  In \nuc{131}{Eu}, where fine structure has been
seen, the $E_{2^+}$ energy is known.  In other cases, systematic trends
must be used.  In actual calculations, we have used the $N_p N_n$
scheme \cite{[Cas85]} to estimate $E_{2^+}$.
  



\section{Numerical Implementation} \label{sec:numeric}

For realistic potentials, the radial Schr\"odinger equation cannot be
solved analytically but must be integrated numerically.  For spherical
potentials, one deals with a single radial equation instead of the
full set (\ref{cc}). In Ref.~\cite{[Ver82]} the code {\sc GAMOW} was
introduced, which uses the Fox-Goodwin method for solving the radial
equation. A more powerful method, the piecewise perturbation, is used
for the same purpose in Ref.~\cite{[Ixa85]}.  The main features are
similar in the two codes.  The total $r$ domain of coupled-channel
equations (\ref{cc}) is separated into two parts.  The first segment
lies along the real axis, $I_1 = [0, r_{\rm max}]$.  The other
interval extends along the complex ray, $I_2 = [r_{\rm max}, r_{\rm
  as}]$, where $r_{\rm as}$ is complex and far enough away that at
$r_{\rm as}$ the asymptotic series of the outgoing Coulomb wave,
$O_l(\eta_{J_d},r_{\rm as} \kjd)$, is a good approximation.  For a resonant
state, the second integration region must be complex for our
regularization scheme given by Eq.~(\ref{sphnorm}).  The rotation
angle of $r$ in $I_2$ should satisfy the condition
\begin{equation}
  \pi - \arg(\kjd) > \arg(r_{\rm as}) > -\arg(\kjd) 
\end{equation}
so that the solution converges along the complex ray.

For axially deformed $V$, the set of coupled-channel
equations (\ref{cc}) must be solved numerically. The piecewise perturbation
method \cite{[Ixa84]} has been generalized for the coupled-channels 
case~\cite{[Ver99]}.
A large value of $r_{\rm as}$  is used, which is far
enough away that the off-diagonal terms of the coupling 
matrix vanish and the asymptotic series for the Coulomb
functions
are accurate.
At this point the coupled-channels equations decouple.  For an initial
$\kjd$ value, one has to calculate the components of a ``left''
solution, $u_{\alpha}^{L}$, which vanish at the origin.  These are
integrated outwards to a matching radius, $r_{\rm m}$, in region $I_1$.
The components of the ``right'' solutions, $u_{\alpha}^{R}$, are
integrated inwards from $r_{\rm as}$ along the complex ray $I_2$.  At
$r_{\rm max}$,  the integration path turns along the real axis to the
matching radius, $r_{\rm m}$.  All components of the ``left'' and
``right'' solutions are linear combinations of linearly independent
solutions of Eq.~(\ref{cc}) in the corresponding $r$ regions.
The two solutions and their derivatives with respect to $r$ should
match at the matching radius and form a set of functions which are
continuous in $r$.  This condition gives a homogeneous set of linear
equations for the unknown expansion coefficients of $u_{\alpha}^{L}$
and $u_{\alpha}^{R}$.  Non-trivial solutions exist only for the
generalized $\kjd$ eigenvalues where the determinant of the set of
linear equations is equal to zero.  For the initial value of  $\kjd$, the
 determinant is not zero; however, it is possible to find
the zero of the determinant by iteration, e.g., using the
Newton--Raphson method.  For the known proton emitters, the width of
the resonance is so small that extremely high numerical accuracy is
needed to calculate the generalized complex energy eigenvalue ${\cal E}$.
We have found that extended precision arithmetic must be employed to
calculate the imaginary part of ${\cal E}$ accurately.  The width
calculated directly in this manner matches well with the current
expression~(\ref{partcur}).


\section{Applications of the Method} \label{sec:phys}
\begin{figure}[thb]
  \begin{center}
\leavevmode
    \epsfig{file=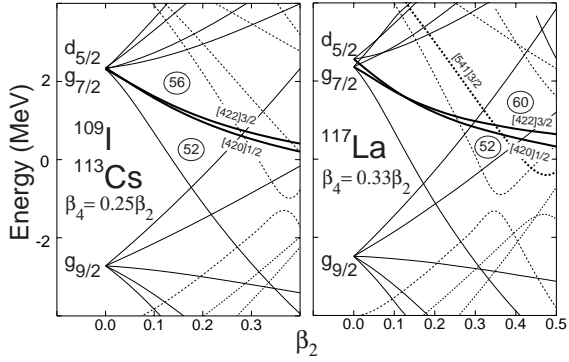,width=8cm}
    \end{center}
\vspace*{-1cm}
  \caption{Single-proton  levels representative of  odd-$Z$ 
    rare-earth nuclei with $Z$$\approx$55 plotted as functions of the
    quadrupole deformation $\beta_2$.  The hexadecapole deformation
    $\beta_4$ was assumed to be proportional to $\beta_2$ to give both
    the spherical and ground-state deformations.  The Nilsson orbitals
    studied for $^{109}$I, $^{113}$Cs, and $^{117}$La are marked by
    thick solid lines.}
    \label{fig:Z=55}
\end{figure}

\begin{figure}[btp]
  \begin{center}
\leavevmode
    \epsfig{file=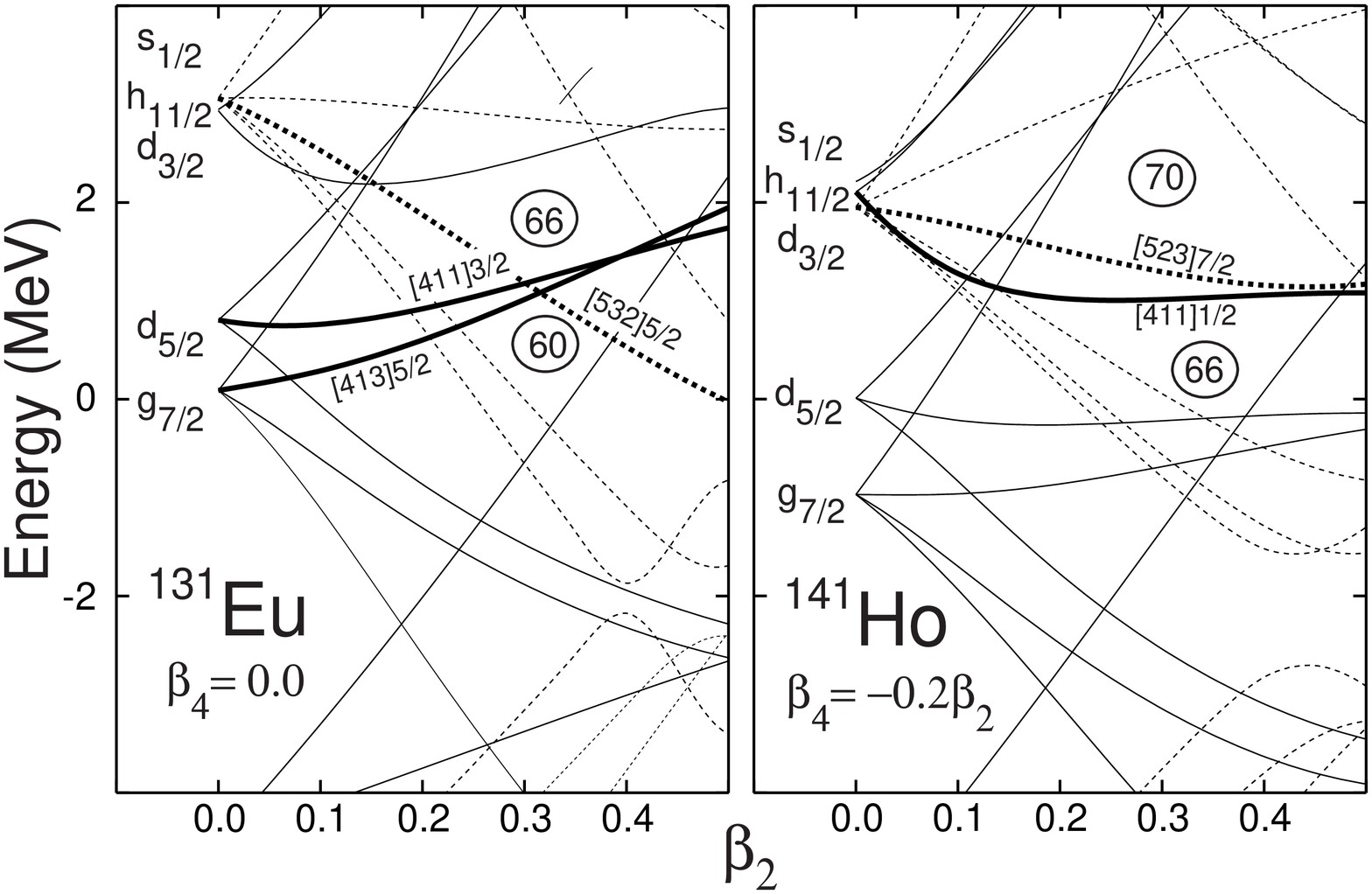,width=8cm}
    \end{center}
\vspace*{-1cm}
  \caption{Same as in Fig.~\protect\ref{fig:Z=55} except for
      \protect\nuc{131}{Eu} and \protect\nuc{141}{Ho}.}
    \label{fig:Z=67}
\end{figure}

This section contains applications of the formalism to measured deformed proton emitters.
For an easy orientation, Figs.~\ref{fig:Z=55} and  \ref{fig:Z=67} show
the proton Nilsson diagrams characteristic of
 $Z$$\approx$55  and $Z$$\approx$67 nuclei, respectively. In our theoretical
 analysis,
 all Nilsson levels close to the Fermi level were investigated.  The
potential depth was always adjusted at each deformation so as to reproduce the
experimental $Q_p$ value.  Table~\ref{tab:Qp} lists the $Q_p$ values,
the energy of the $2^+$ states, and deformation parameters for all the 
nuclei investigated in this paper.

\begin{table}[tb]
  \begin{tabular}{rllcc}
    & $Q_p$ (keV) & $E_{2^+}$ (keV) & $\beta_2$ & $\beta_4$ \\ \hline \\
    \nuc{109}{I}   & 829(4)~\cite{[Sel93]}  &  625~\cite{[Dom94]} &
    0.09  & 0.03 \\
    \nuc{113}{Cs}  & 977(4)~\cite{[Pag94a]}  &  466~\cite{[Smi00]}   &
    0.16  & 0.04 \\
    \nuc{117}{La}  & 800(10)~\cite{[Sor99]} & 150   & 0.30 & 0.10 \\
    \nuc{131}{Eu}  & 950(7)~\cite{[Son99]} & 121(3)~\cite{[Son99]} &
    0.32 & 0.00 \\
    \nuc{141}{Ho}  & 1.190(10)~\cite{[Ryk99]} & 160 & 0.29 & -0.06
    \\
    \nuc{141m}{Ho} & 1.251(20)~\cite{[Ryk99]} & 160 & 0.29 & -0.06
  \end{tabular}
  \caption{List of $Q_p$ values, $2^+$ state energies, and deformation
    parameters for nuclei investigated.  $E_{2^+}$ energies without a
    reference were estimated using the $N_p N_n$ scheme of
    Ref.~\protect\cite{[Cas85]}.}
  \label{tab:Qp}
\end{table}

\subsection{Description of Rotational Bands Built Upon Deformed Resonances}
As has been previously mentioned, a significant benefit of working in
the nonadiabatic formalism is the proper treatment of the
ground-state rotational band in the daughter nucleus.  This makes it possible
to easily calculate  the fine structure in the proton emission.
The presence of  the rotational band in
the daughter nucleus also gives rise to rotational bands built upon
$J=\Omega$ bandheads in the parent nucleus.  
 In a previous work~\cite{[Kru00]}, we discussed a rotational
band in \nuc{131}{Eu} built upon the $J=\half{3}^+$ level associated
with the \nil{411}{3} Nilsson orbital.  The spacing of the levels in the
parent nucleus follow nicely the expected $J(J+1)$ spacing with the same
moment of inertia parameter as assumed for the daughter nucleus. 
Small deviations from the $J(J+1)$ spacing result from the Coriolis
coupling.

To verify that 
the calculated band structure indeed belongs to the same {\em intrinsic} Nilsson configuration,
one  can inspect
the $K$-decomposition of each rotational level.  This is done by using
Eq.~(\ref{adiabatic}) to project the nonadiabatic wave functions 
onto adiabatic states with good $K$.  For the $J$=$\half{3}^+$
orbital in \nuc{131}{Eu}, the $K$-decomposition is
shown in Table~\ref{tab:k-dec}.  
It is seen that the $K$=$\half{3}$ dominates, although there appear small
admixtures of other
$K$ components due to the Coriolis coupling. Note the presence of the 
$K$=$\half{1}$ which is forbidden in the strong coupling limit.
 
\begin{table}[tb]
    \begin{tabular}{lllll}
      spin($J$) & $K=\half{1}$ & $K=\half{3}$ & $K=\half{5}$ &
      $K=\half{7}$ \\ \hline 
      $3/2$ & $0.0017$ & $0.9972$ & &  \\
      $5/2$ & $0.0040$ & $0.9894$ & $0.0056$ &  \\
      $7/2$ & $0.0095$ & $0.9770$ & $0.0013$ & $1.77 \times 10^{-5}$ 
    \end{tabular}
    \caption{$K$-decomposition of the calculated band members of the \nil{411}{3} 
     band  in \nuc{131}{Eu}.  The $K \neq \half{3}$
      components arise from  the Coriolis coupling.}
    \label{tab:k-dec}
\end{table}

\begin{figure}[htb]
 \begin{center}
\leavevmode
   \epsfig{file=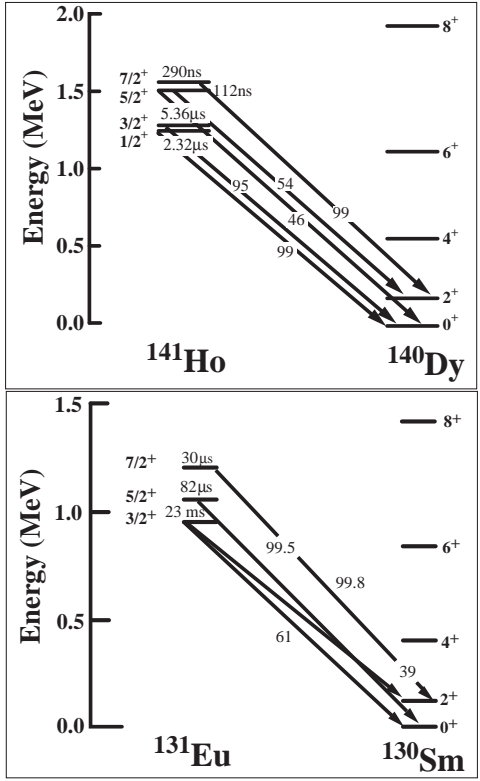,width=8cm}
 \end{center}  
    \caption{Rotational bands in 
     \nuc{131}{Eu} (bottom; built upon the [411]3/2 Nilsson level) and
      \nuc{141}{Ho} (top; built upon the [411]1/2 Nilsson level). 
  For \nuc{131}{Eu}, the strongly coupled rotational band is predicted.
 For \nuc{141}{Ho},  the two decoupled, almost degenerate, sequences
 are calculated.
 Proton lifetimes and strongest branching ratios are indicated.} 
    \label{fig:decouple}
\end{figure}

A very different picture arises for the $J$=$\half{1}^+$ band built upon
the \nil{411}{1} Nilsson orbital in \nuc{141}{Ho}.  Its low-lying 
band members,
through $J$=$\half{7}$,  are shown in
Fig.~\ref{fig:decouple}.  In this case, we do not see the
development of a strongly coupled  band as in \nuc{131}{Eu}, but rather
two nearly degenerate decoupled signature partners.  This comes about due to
the large decoupling parameter for this orbital.  Since \nuc{141}{Ho}
is well deformed, we can consider the Coriolis interaction as a
perturbation in the strong coupling approximation.  For a $K=\half{1}$
band, first-order perturbation theory gives \cite{[Nil55]}
\begin{equation}
  \label{eq:decoupling}
  E^{J}_{\half{1}} = E^{0}_{\half{1}} + \frac{1}{2{\cal J}} \left\{
      J(J+1) - \frac{1}{4} + a_d (-)^{J+\half{1}}(J+\half{1})
    \right\},
\end{equation}
where $a_d$ is the decoupling parameter.  For a non-zero decoupling
parameter, the $J+\half{1}$ odd levels are shifted against the
$J+\half{1}$ even levels with a degeneracy setting in for $|a_d| =1$.
From studies of well-deformed and super-deformed bands in odd-$Z$ rare-earth nuclei, 
bands built on the \nil{411}{1} level are known to have 
 a decoupling parameter near
$-1$~\cite{[Naz90a],[Bak95]}.  This nicely explains our predictions and gives yet
another verification that  the weak coupling
formalism properly
incorporates the Coriolis interaction.
It needs to be noted that the branchings shown in
Fig.~\ref{fig:decouple} correspond to the proton emission only. 
In reality, the low-lying levels in these bands
rapidly decay by
gamma radiation ($\Gamma_{\gamma} \gg \Gamma_p$); that is,
the lifetimes of these states are much shorter.

It is interesting to look in detail at the make-up of the cluster
radial wave function and the partial widths.  In the partial wave
decomposition, the dominant components are those of the originating
spherical state.  For example, in \nuc{117}{La}, the \nil{422}{3}
Nilsson orbital originates from a $g_{7/2}$ spherical state.  At a
deformation of $\beta_2 = 0.33$, the wave function still contains
$60\%$ of $g_{7/2}$ distributed between the $2^+$ and $4^+$ daughter
states.  However, due to deformation,
other partial waves with $j \geq \half{3}$ also contribute:
$d_{3/2}\,(9.6\%)$, 
$d_{5/2}\,(9.0\%)$, $g_{9/2}\,(14.2\%)$, and $i_{11/2}\,(3.5\%)$.  
Coriolis coupling
introduces the $s_{1/2}$ partial wave with an amplitude of $4.2\%$.
Although the radial wave function is a  combination of  components having
different angular momentum, the decay branches are easy to understand.  The
total width is governed  by the high penetrability of low-$l$ partial waves.
In fact, $97\%$ of the width of this resonance is in the
$d_{3/2} \otimes 0^+$ channel.  The remaining part comes from the $d_{3/2} \otimes
2^+ (0.6\%)$ and the $s_{1/2} \otimes 2^+ (2.3\%)$ channels.  

The majority of decays investigated in this work have small branching
ratios, less than ten percent.  However, a few have quite large
branching ratios to $2^+$ states, including the possible decay out of the \nil{532}{5}
Nilsson orbital in \nuc{131}{Eu} which is predicted in this work to have the  
branching ratio of 52\%.
The circumstances that lead to such large branching ratios are worthy
of investigation.  The \nil{532}{5} orbital originates from an
$h_{11/2}$ spherical orbital.  At a  deformation of $\beta_2=0.32$,
the \nil{532}{5} orbital consists mainly of $h_{11/2} (75\%)$,
$f_{7/2} (18\%)$ and only $1.9\%$ of $f_{5/2}$.  There is an
additional $0.8\%$ of the $K$-forbidden $p_{3/2}$ component.  The
decay to the ground state can proceed only via the $f_{5/2}$
component.  Meanwhile, the decay to the $2^+$ state proceeds mainly through
the $p_{3/2}$ and $f_{7/2}$ waves; the former due to the lower
angular momentum and the latter due to the larger make-up in the total
wave function.  The combination of a low-lying excited state, a lower
angular momentum channel, and suppressed amplitude of the $f_{5/2}$ wave 
leads to the very high branching ratio this state would exhibit.

\subsection{Branching Ratios}
The main impetus behind this work has been the recently observed
fine structure in the proton decay of \nuc{131}{Eu}~\cite{[Son99]}.
The nonadiabatic formalism offers great advantages over the
strong-coupling approximation in calculating fine structure.  The
proper placement of the daughter states are explicitly included and
the channels are now labeled with the proton's orbital and total
angular momentum, $lj$, and the angular momentum of the daughter
nucleus, $J_d$.  In one fell swoop, both the lifetime and partial
widths are calculated.

\begin{figure}[bt]
  \begin{center}
  \leavevmode
    \epsfig{file=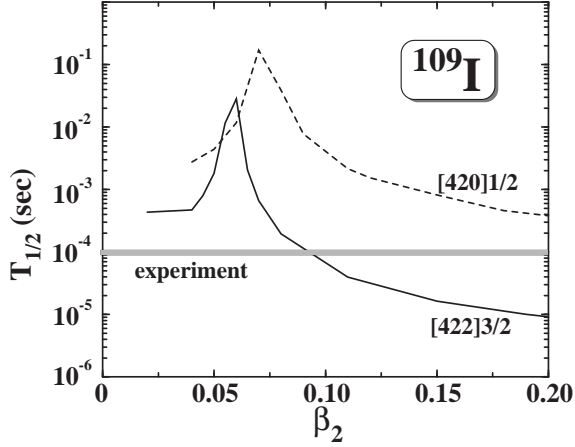,width=8cm}
 \end{center}
    \caption{Predicted lifetimes in \nuc{109}{I} for the [422]3/2 and [420]1/2 orbitals.
     The experimental
      lifetime is $110(5) \mu$s~\protect\cite{[Sel93]}.}
    \label{fig:109I}
\end{figure}

As was shown previously in Refs.~\cite{[Ryk99],[Kru00]}, for large
deformations our calculations show little sensitivity to $\beta_2$ and
$\beta_4$.  This is because the spherical decomposition of the
corresponding Nilsson orbitals varies little in this regime, provided
that there are no crossings between the Nilsson orbitals of interest.
The uncertainty due to nuclear deformation is usually smaller than
that due to experimental uncertainty in the proton energy.  In the
less-deformed cases, there is a greater dependence of $\beta_2$ and
$\beta_4$.  In the $1/2^+$ level in \nuc{109}{I}, shown in
Fig.~\ref{fig:109I}, we see the effect of a level crossing near
$\beta_2 \approx 0.05$.  (This effect has been noted earlier in
Ref~\cite{[Mag99]}.)

\begin{table}[htb]
    \begin{tabular}{llccr}
  &  Orbital &  $u^{2}$ & $\tau_{1/2}$ &
    b.r. \\ \hline \\
  \nuc{109}{I} & \nil{420}{1}  & 0.99 & 94.8 $\mu$s & 0\% \\
               & \nil{422}{3}  & 0.99 & 7.86 ms     & 0\% \\  
               &               &      &\boldmath{$110(5) \mu
                 s$}~\cite{[Sel93]}& \\  \\

  \nuc{113}{Cs}& \nil{420}{1}  & 0.52 & 0.66 $\mu$s   &  0\%  \\
               & $J=3/2^{+}$   & 0.56 & 34.7 $\mu$s   &  0\%  \\  
               &               &      &\boldmath{$16.7(7) \mu
                 s$}~\cite{[Bat98]}&   \\  \\

  \nuc{117}{La}& \nil{420}{1}  & 0.32 & 1.27 ms     & 0\% \\
               & \nil{422}{3}  & 0.33 & 103 ms      & 3\% \\
               & \nil{541}{3}  & 0.61 & 293 ms      & 4\% \\
               &               &      &{\bf 20(5)
                 ms}~\cite{[Sor99]} &  \\ \\

  \nuc{131}{Eu}& \nil{411}{3}  & 0.71 & 34.0 ms     & 39\% \\
               & \nil{413}{5}  & 0.52 & 184  ms     &  7\% \\
               & \nil{532}{5}  & 0.48 & 3.90 s      & 52\%  \\
               &              &   & {\bf 17.8(19) ms} & 
                                {\bf 24(5)\%}~\cite{[Son99]}  \\ \\

  \nuc{141}{Ho}&\nil{411}{1}   & 0.70 & 14.6 $\mu$s & 0.8\% \\
               &\nil{523}{7}   & 0.84 & 19.1 ms     & 6\%   \\
               &               &      & {\bf 3.9(5)
                 ms}~\cite{[Ryk99]}  & \\ \\

  \nuc{141m}{Ho}&\nil{411}{1}  & 0.70 & 3.3 $\mu$s  & 1\% \\
                &\nil{523}{7}  & 0.84 & 4.6 ms      & 9\% \\
                &              &      &\boldmath{$8(3) \mu
                  s$}~\cite{[Ryk99]} & 
    \end{tabular}
    \caption{Table showing the various orbitals for each system
      investigated in this work.  Except for the weakly deformed
      systems of \nuc{109}{I} and \nuc{113}{Cs}, the deformation
      dependence is much weaker than the uncertainty due to the
      experimental $Q_p$ value.  The theoretical spectroscopic
      factor, half-life, and branching ratio to the $2^+$ states are
      shown.  Experimental results (where available) are shown in bold
      type. }
    \label{tab:results}
\end{table}

Table~\ref{tab:results} shows predicted half-lives, theoretical
spectroscopic factors, and branching ratios.  The spectroscopic factors 
have been estimated in the independent-quasi-particle picture.  Note
that the $1/(\Omega+\half{1})$ factor present in  the strong-coupling
approximation is no longer needed.  Our predictions for \nuc{131}{Eu}
and the ground and isomeric states in \nuc{141}{Ho} are unchanged from 
\cite{[Kru00]}.  The ground state of \nuc{131}{Eu} is consistent with
the \nil{411}{3} assignment.  This is the same conclusion as in
Refs.~\cite{[Son99],[Mag00]} but differs from the assignment 
of \nil{413}{5} of Ref.~\cite{[Mag99]}.  

In  \nuc{141}{Ho}, the assignments are straightforward: \nil{523}{7} for the
ground state and \nil{411}{1} for the isomeric state.  These match the 
assignments of Refs.~\cite{[Dav98],[Mag99]}.  In \nuc{109}{I} we find
agreement with the \nil{420}{1} with a deformation near $\beta_2$=0.10.
  This agrees with suggestions of Refs.~\cite{[Bug89],[Mag99]}.  
In \nuc{113}{Cs} we see a large admixture of $K$=1/2 in the $J$=3/2
wave function.  Therefore, the asymptotic
Nilsson labeling  is inappropriate, and
only the total angular momentum is used to label the state in
Table~\ref{tab:results}.  The two orbitals near the Fermi level in
\nuc{113}{Cs} correspond  to a pseudospin doublet; hence, strong Coriolis 
mixing of these levels.
In the newly
discovered proton emitter \nuc{117}{La}, the experimental lifetime is
20(5) ms~\cite{[Sor99]}.  It appears that the \nil{422}{3} assignment
is best with a lifetime of 100 ms at a deformation of $\beta_2$=0.30, 
$\beta_4$=0.11.  

\begin{figure}[tb]
  \begin{center}
 \leavevmode
    \epsfig{file=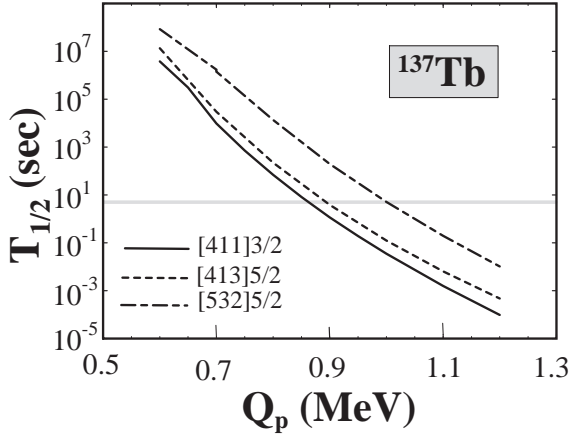,width=8cm}  
  \end{center}
  \caption{Predicted half-lives for \nuc{137}{Tb} as a function of
    proton $Q_p$ value.  The assumed deformation is $\beta_2=0.28$ and
    the estimated value of $E_{2^+}$ is $165$ keV.  This figure is
    meant to identify the regions of $Q_p$ and $T_{1/2}$ to look for
    this yet-unseen proton emitter.  For longer lifetimes, serious
    competition from beta decay is expected.  Above the grey line, the
    branching ratio for proton decay is predicted to be less than
    $10\%$~\protect\cite{[Mol97]}. }
    \label{fig:Tb}
\end{figure}

There is currently a proposal~\cite{[Ryk00]} to search for
proton emission from \nuc{137}{Tb}.  Being in the region between
\nuc{131}{Eu} and \nuc{141}{Ho}, this nucleus is expected to be well deformed
with  $\beta_2$$\approx$0.28.  Using the Grodzins
formula~\cite{[Gro62],[Ste72]}, we estimate the energy of the $2^+$
state in \nuc{136}{Gd} to be $165$ keV.  Figure~\ref{fig:Tb} shows the 
expected half-life as a function of $Q_p$.  It is
expected that for lifetimes longer than the limit marked by 
 the grey line, beta-decay will dominate~\cite{[Ryk00]}.

\subsection{Theoretical Uncertainties}
It should be emphasized  that our  method contains no
\emph{adjustable} parameters; there are a few parameters which are
set by experiment.  These include $Q_p$ and the placement of the 
lowest few levels in the ground-state band of the daughter nucleus.
Since the higher levels are energetically forbidden, even if they are
needed in the calculation to ensure proper convergence, the
half-lives and branching ratios are fairly insensitive to their placement.
We shall now discuss the sensitivity of the calculated half-lives and
branching ratios to various quantities used in the calculations.  For
concreteness, we will focus on the \nil{411}{3} level in
\nuc{131}{Eu}.  All other levels studied show similar sensitivities.

The largest effect on the lifetime comes from the $Q_p$ value.  The
$Q_p$ value for  \nuc{131}{Eu} is currently taken as 950(7)
keV~\cite{[Son99]}.
The uncertainty of 7 keV leads to an uncertainty in the calculated
lifetime of $-7.5/+9.8$ ms.  This is a difference of roughly $-22/+30
\%$.  Since a change in the $Q_p$ value also affects the energies of excited
states, the change in branching ratio is much smaller.  For the
\nil{411}{3} orbital, the effect is $\pm 1.3$ \%.  

On the other hand, the placement of the $2^+$ level has  a smaller
effect on the lifetime but greatly influences the branching ratio.
Based on  Ref.~\cite{[Son99]}, the $2^+$ level in \nuc{130}{Sm} 
is placed  at
121(7) keV. This 7 keV uncertainty changes the lifetime by $\pm 4.0$
ms ($\pm 12\%$).  For the branching ratio, the corresponding error  is $\pm 6.7$\%.

In the nuclei with significant branching ratios, little to nothing is
known about the level structure in the daughter system; hence, we had
to assume a perfect rotor to assign energies to the states above the
$2^+$.  To check for the sensitivity to this assumption, we repeated
some calculations assuming $E_{J_d} = \kappa'J_d(J_d+1) -
B[J_d(J_d+1)]^2$.  The anharmonicity factor, $B$, has typical values
around $\kappa'/200 \approx 100$ eV~\cite{[Szy83]}.  This introduces a
1.0 ms shortening of the lifetime and a reduction of the branching
ratio of 1.2\%.  Both are much smaller than the influence of the
$Q_p$ value or $E_{2^+}$.  So as long as the proper $Q_p$ value is
used along with a good estimate of the first excited state, the
remaining part of the spectrum needs only to be reasonably placed.

Additional uncertainties can arise from the optical model potentials.
As discussed in Sec.~\ref{sec:mod-para}, we believe that the Chepurnov 
parameterization is the best current compromise.  (It is noted here that
better agreement between theory and experiment could,
in principle, be achieved by
fitting the optical model parameters to the properties, including
proton decay data, of these drip-line nuclei.)  As discussed in
Ref.~\cite{[Abe97]}, the lifetime of spherical proton emitters depends
weakly on the nuclear structure details.  Reasonable variations in
radius and diffuseness parameters affect the lifetimes by less than a
factor of about three.


\section{Assessment of the adiabatic approximation} \label{sec:adi}

As discussed in Sec~\ref{sec:theory}, all previous work
on deformed proton emitters have made the adiabatic approximation (AD)
\cite{[Dav98],[Ryk99],[Son99],[Bug89],[Fer97],[Mag98],[Mag99],[Tal98],[Fer00],[Mag00]}.
The use of the nonadiabatic formalism for proton emission was first
used by us in the recent Ref.~\cite{[Kru00]}.  The power of the nonadiabatic
approach is apparent in several areas.  First, due to the fact that
the wave function is in the laboratory frame, the Coriolis coupling is
implicitly included.  This allows for the inclusion of all the  partial
waves with $j < J$ in the proton's wave function.  In particular, 
the Coriolis coupling can admix
states with smaller $l$-values, and consequently lower centrifugal
barriers, into the proton wave function.  Low-$l$ components, however small,
can substantially affect the lifetime.
\begin{figure}[tb]
\begin{center}
 \leavevmode
  \epsfig{file=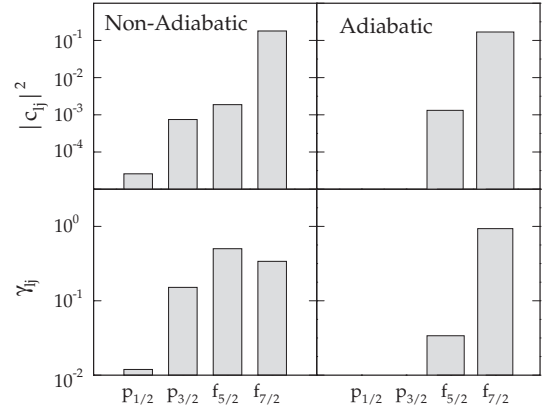,width=8cm}
\end{center}
  \caption{Comparison of partial widths and wave function amplitudes
    obtained in the nonadiabatic (left) and adiabatic (right)
    calculations for the \nil{532}{5} deformed resonance in
    \nuc{131}{Eu}.  Only the lowest few partial waves are shown. The
    upper panels show the spherical amplitudes $|c_{\alpha}|^2$.
    Notice the presence of the $K$-forbidden $p$-wave components in
    the nonadiabatic approach.  The lower panel shows the normalized
    partial widths $\gamma_{\alpha} \equiv \Gamma_{\alpha} /
    \Gamma_{\rm tot}$. In the non-adiabatic case we have summed over
    all possible daughter states.}
  \label{fig:cori}
\end{figure}
Figure~\ref{fig:cori} shows this effect for the \nil{532}{5} deformed
resonance in \nuc{131}{Eu}.  Note that the $p_{3/2}$ partial wave
contributes only $0.3\%$ to the total wave function, yet accounts for
$15\%$ of the decay width.

In order to calculate the branching ratio in the adiabatic
approximation, some ansatz must be used.  Firstly, the partial width
to the ground state is approximated by the width for the partial wave
that matches the initial state,
\begin{equation}\label{eq:adi_gs_width}
  \Gamma_{0^+}^{\rm ad} = \frac{1}{\Omega + \half{1}}
  \Gamma_{lj=\Omega}^{\rm ad} .
\end{equation}
For the excited states, a weighted sum over the possible partial
wave components is used:
\begin{equation}\label{eq:adi_es_width}
  \Gamma_{J_d}^{\rm ad} = \frac{2J_d + 1}{\Omega + \half{1}} 
  \sum_{lj} \, |\langle J_d 0 \; j \Omega | \Omega\Omega \rangle |^2 \;\;
  \Gamma_{lj}^{\rm ad}(Q_{J_d}) ,
\end{equation}
where the partial widths have been calculated with a $Q_p$ value
adjusted to that of the $J_d^+$ state.  This  reduces to
Eq.~(\ref{eq:adi_gs_width}) for $J_d = 0$.  This procedure
will be referred to as the adiabatic corrected method (ADC).

\begin{table}[tb]
  \begin{tabular}{llccc}
    Nucleus & Orbital & $\tau_{\half{1}}$ (AD) & $\tau_{\half{1}}$
    (ADC) &  b.r. \\
    \hline\\
    \nuc{109}{I}  & \nil{420}{1} & $9.85 \times 10^{-6}$  &
    $24.1\times 10^{-6}$  &  0\%  \\
    \nuc{113}{Cs} &  $J=3/2^{+}$ & $1.98\times 10^{-6}$  &
    $7.02\times 10^{-6}$  &  0\%  \\
    \nuc{117}{La} & \nil{422}{3} & $20.5\times 10^{-3}$  &
    $90.3\times 10^{-3}$  &  0.3\%  \\
    \nuc{131}{Eu} & \nil{411}{3} & $868.\times 10^{-6}$  &
    $48.7\times 10^{-3}$  &  37\%  \\
    \nuc{141}{Ho} & \nil{523}{7} & $5.95\times 10^{-3}$  &
    $6.66\times 10^{-3}$  &  3\%  \\
    \nuc{141m}{Ho}& \nil{411}{1} & $1.94\times 10^{-6}$  &
    $3.43\times 10^{-6}$  &   1\%  
  \end{tabular}
  \caption{A calculation of the proton lifetimes using the adiabatic
    formalism of Sec.~\protect\ref{sec:strong}.  All lifetimes are in seconds.
    The label (AD) corresponds to calculations in the
    adiabatic approximation, i.e., it includes all degenerate final states.
    The column ADC corresponds 
to the  corrected adiabatic approximation, in which 
it is assumed that  the decay goes only to the
    ground state of the daughter nucleus,
    Eq.~(\protect\ref{eq:adi_gs_width}).  The branching ratio is
    calculated using Eqs.~(\protect\ref{eq:adi_gs_width}) and
   (\ref{eq:adi_es_width}). For the nonadiabatic model
predictions, cf. Table~\ref{tab:results}.  } 
  \label{tab:adi}
\end{table}

As can be seen in Table~\ref{tab:adi}, in some instances the adiabatic 
and  nonadiabatic predictions are very close, like for \nuc{141m}{Ho}
and \nuc{117}{La}.  However, 
in other cases the differences are striking, like
the factor of five difference in \nuc{113}{Cs}.  In those systems
where the agreement is good, there is no admixture of lower-$l$
partial waves in the nonadiabatic formalism.  

It is also worth noting that lifetimes calculated in the full
adiabatic method are usually shorter by a factor of up to four as
compared to the ADC method.  However, for the \nil{411}{3}
orbital of \nuc{131}{Eu}, there is a factor of 56 difference.  This
results from the large $J_d^\pi$=$2^+$ component in the corresponding
Nilsson model function.

In about half of the cases studied, the adiabatic approximation,
particularly with angular momentum conservation enforced by hand,
gives results similar to the nonadiabatic method.  In the rest, the
difference can be large.


\section{Conclusions} \label{sec:conc}

The state-of-the-art coupled-channel formalism has been extended to
include excitation modes in the daughter system.    The weak-coupling scheme
applied  allows us to work in the laboratory reference frame.
The exact treatment of
excitation spectrum in the daughter nucleus also allows a consistent
calculation of branching ratios.  

As could be expected, significant branching ratios are expected only
for well-deformed nuclei where the first excited state of the daughter
nucleus lies low in energy.  The Coriolis mixing of states with lower
orbital angular momentum can enhance the decay to the excited state,
e.g., the decay of the \nil{532}{5} orbital in \nuc{131}{Eu} where the
branching ratio to the $2^+$ state is predicted to be as large as
$52\%$.

In the case of spherical proton emitters, the proton separation
energy and orbital angular momentum have the largest effect on
lifetimes~\cite{[Abe97]}.  The detailed nuclear structure plays a
minor role.  In the deformed case, the placement of the
single-particle levels also has a significant effect.  This is true in 
both the adiabatic formalism, where mixing occurs with higher-$l$
states, and in the nonadiabatic formalism where the Coriolis coupling 
can mix states with lower-$l$ also.  

We have been able to calculate the placement and decay properties of
excited levels in the ground-state band of the parent nucleus.  As
shown in Fig.~\ref{fig:decouple}, both strongly coupled  
and nearly degenerate decoupled  bands are
predicted, depending on the nature of the band head.
 While we have calculated the proton decay half-lives of the excited band members,
they are all much too slow to compete with in-band gamma-decay.
In all cases investigated, by comparing theoretical predictions  with
experimental half-lives and branching ratios (where available), we have 
been able to identify the Nilsson orbital which the proton occupies.
We have also been able to discern the angular momentum components of
the proton wave function.  

While our calculations for well-deformed nuclei give a quantitative
agreement with experiment, it would not be proper to apply the present
model to vibrational or transitional nuclei such as \nuc{151}{La}. For
this, one needs to redefine the coupling matrix elements in
Eq.~(\ref{Vaa'}). Calculations along these lines are in progress.

\acknowledgments
Useful discussions with Krzysztof Rykaczewski are gratefully acknowledged.
This research was supported in part by
the U.S. Department of Energy
under Contract Nos.\ DE-FG02-96ER40963 (University of Tennessee),
DE-FG05-87ER40361 (Joint Institute for Heavy Ion Research),
and
DE-AC05-00OR22725 with UT-Battelle, LLC (Oak
Ridge National Laboratory),
Hungarian OTKA Grants No. T026244 and No. T029003, and UNISOR
(UNISOR is a consortium of universities supported by its members and by
the U.S. Department of Energy.)



\end{document}